\documentclass[10pt, oneside, a4paper, twocolumn]{article}

\usepackage[cjk]{kotex}
\usepackage{setspace}
\usepackage[a4paper,left=20mm,right=20mm,top=30mm,bottom=30mm]{geometry}
\usepackage{tikz}
\usepackage[letterspace=200]{microtype}
\usepackage{graphicx}
\usepackage{amssymb}
\usepackage{endnotes}
\usepackage{amsmath}
\usepackage[normalem]{ulem}
\useunder{\uline}{\ul}{}

\graphicspath{ {fig/} }

\usepackage{fancyhdr}
 
\begin{document}

\onecolumn
\setstretch{1.2}

\title{Market Crash Prediction Model for Markets in A Rational Bubble}

\author{HyeonJun Kim\medskip\\{\normalsize UG in Dept.Finance, Soongsil University}
\medskip\\{\normalsize hyeonjunacademic@gmail.com}}


\maketitle

\begin{abstract}
   
Renowned method of log-periodic power law(LPPL) is one of the few ways that a financial market crash could be predicted. Alongside with LPPL, this paper propose a novel method of stock market crash using white box model derived from simple assumptions about the state of rational bubble. By applying this model to Dow Jones Index and Bitcoin market price data, it is shown that the model successfully predicts some major crashes of both markets, implying the high sensitivity and generalization abilities of the model. \\\\

\textbf{keywords:} {[Rational Bubble], [Market Crash]}


\end{abstract}

\setstretch{1.5}



\section{Introduction}

It is well known that the efficient market price tends to follow the characteristics of Brownian Motion that strictly depends on stochastic nature\cite{merton1975optimum}. Thus predicting the information about the short-term return of a financial portfolio is known to be difficult, but there are few exceptions. Firstly, there is a widely held belief among traders that it is possible for some computational deterministic patterns to emerge for short period of time. This belief, combined with econophysics, led the creation of algorithmic trading funds such as Prediction Company and Renaissance Technologies. Secondly, there is a concept of rational bubble that illustrates the rational behavior of people who invest in markets that clearly creating an economic bubble\cite{diba1988theory}. 

With these two facts in mind, it is possible to imagine a model with reasonable assumptions to predict rational bubble's key behaviors, such as the stock market crash after a bubble economy.
While there are more sophisticated way of prediction such as log-periodic power law, there seems to be no clear white-box model that also have high prediction abilities. In this paper, by using some reasonable assumptions about the investors, a white-box model about the stock market state will be derived. The model will then be applied to two different cases, the Dow Jones Index and Bitcoin market prices to assess the ability of the model.

\subsection{Previous Study on Rational Bubble and its Predictions}

Theories and methodologies of complex systems helped studies and individuals to make attempts to solve the prediction problems. Log-periodic power law(LPPL) is one of them\cite{sornette2009stock}, illustrating the dynamics of two conflicting group of investors, value investors and trend followers\cite{ide2002oscillatory}. There are some successful Implementation cases of this model as well, demonstated by Yan, Woodarda, Sornette\cite{yan2010diagnosis}.

\section{Model and its Assumptions}

\subsection{Definition of Rational Bubble}

A rational bubble in this paper's context is the state of a market that the majority of the new investors are investing because of the past market price or any major index. For example, if investors of NASDAQ invest in NASDAQ only because the past NASDAQ index was rising, it is called a rational bubble in this paper. There is more formerly known definition about this particular word. However for the convenience of understanding the model's key points, the definition will be little vague for now.

\subsection{Assumptions}

Because investing in markets with rational bubble is caused by the price change, these investors are only interested in price or similar indices. This causes the feedback loop in the market, as a self-fulfilling enthusiasm caused by the investors themselves\cite{yan2010diagnosis}. There, if the price point $p_\alpha$ where average return of the investors is same as the market return is higher than the current market price point, it is somewhat true that most investors' portfolio is showing less return than the market presents. This is because not all investors invested in the lowest price point, making the average return of the portfolios grow slower than the market return during the bubble even though the market price change is same to every investor. Thus, if the market price exceeds price point $p_\alpha$, the investors expectation for return caused by the superficial market return is not satisfied. This makes the investors behave more irrationally, or doubt the market, as the root assumption of their investment is not met. This potential tendency could cause massive market crash just with somewhat minor price change, causing the doubt to become a real action. All of this potential threat is possible after the minimum price point where market return is same as the average return of the investors.

\subsection{Mathematical Derivation of the Model}

For convenience, it is assumed that every investors invest in the market portfolio. If a certain price point is $p_x$ and the price before the bubble is $p_0$ and assuming that investors of the bubble market started investing after the start of the bubble, the average return at price point $p_x$ of every single stock in a continuous form is as follows.

$$E(R) = {\int_{p_0}^{p_x} R_p V_p \, d p \over \int_{p_0}^{p_x} V_p \, d p} $$

$R_p$ is the current return of a market portfolio in interest which is bought at price $p$, and $V_p$ is the volume of a market portfolio in interest which is bought at price $p$.
As for the volume, the trading volume will be used for approximation. Also, as the exact data of trading volume could not be used because the discrete nature of trading volume, the trading volume will be approximated by a regression line which shows the relationship between price and trading volume. The formula of the regression line is as follows.

$$v = a p + b = {r_{v p} S(v)\over S(p)}  p + \left(E(v) - {r S(v)\over S(p)} E(p)\right)$$

$v$ is volume of trades, $p$ is the price, $S(v)$ and $S(p)$ are the standard deviation of $v$ and $p$,  $E(v)$ and $E(p)$ are the average of $v$ and $p$, and $r_{vp}$ is the correlation coefficient of the two variable $v$ and $p$. By using this regression line, the model could be simplified to the following form.

$$E(R) = {\int_{p_0}^{p_x} {p_x \over p} (a p +b) \, d p \over\int_{p_0}^{p_x} ap + b \, d p } = {a p_x(p_x - p_0 ) + p_x b \ln \left({p_x\over p_0}\right) \over 0.5 a p_x^2 - 0.5 a p_0^2 + b(p_x - p_0)}$$

The resulting average return is in $1+R$ format. As the stability of the market is removed when the average return of the investors becomes substantially less than the market return during the bubble, The minimal price point where market becomes unstable will be the price where market return $R_M(p_x)$ is
same as the average return $E(R)$. In other words, we are finding for $p_\alpha$ where

$$M_R(p_\alpha) = {p_\alpha \over p_0} = E(R) ={a p_\alpha(p_\alpha - p_0 ) + p_\alpha b \ln \left({p_\alpha\over p_0}\right) \over 0.5 a p_\alpha^2 - 0.5 a p_0^2 + b(p_\alpha - p_0)} .$$Lastly, the average return $E(R)$ is compared with the current market return $M_R$ for further evaluation.

\section{Practical Implementation}

As for any model, practical implementation is as important as its thorough mathematical structure. Designing an implementation trial for this model is quite difficult task. Because the model is only designed for assessing markets in some form of the rational bubble, it is hard to determine the sample of the market data that fits the terminology "rational bubble". Also, the trial should be capable of assessing specificity as well. To satisfy all of this requirements, testing for every possible time span of 50 trading days had been considered to be the best option. The model's performance is assessed visually. Also, an implementation trial include non-traditional markets as well, for testing the generalization capabilities of the model. 

Unlike the clarity of the model's assumptions and structure, most of the model's components are not set for practical usage. Therefore, some carefully designed methodology should be used for implementation of the model. Although the specific methodology that been used for this paper's trial is confidential, some of the key problems will be introduced, and the trial's result will be shown and discussed. 

\subsection{Data Preparation}

For the stock market data, the Dow Jones Index from January 1999 to August of 2021 is been used, and for the non-traditional market data, the Bitcoin market price data from January 2016 to August of 2021 is been used.

\subsection{Problems with Implementation}

There are some problems to solve before applying the model to the real world data. First, the regression line between trade volume and price should be calculated without problematic issues like assigning discrete data of volume to the continuous interval of price. Also, as the model includes logarithm, some solvers are incapable of solving the equation $M_R(p_\alpha) = E(R)$ for $p_\alpha$ .

However these problems could be solved by simple practical solutions. The method that used for this trial is confidential. 

\subsection{Implementation Results and Discussion}

When applied to Dow Jones, the model showed significant prediction about 2008 financial crisis and 2020 stock market crash, which are located at around index 2100 and 5300. In Other cases, the model come out with many contradicting results during short time span or show weak predictions. However, the model seems to consider the time span mid 2020 to mid 2021 as the precursor of another market crash. This might means that the model's specificity is lower than expected, or just the error occurred while dealing with wider price gap. Other explanation is that the stock market is really in the rational bubble, but the bubble is sustained by the stimulus plan of US government.
Also, the model seems to also consider the short time span after the market crash as the market crash itself. However this error can be easily adjusted by human evaluator.

\begin{center}

\begin{figure}[h]

\includegraphics[width=16cm]{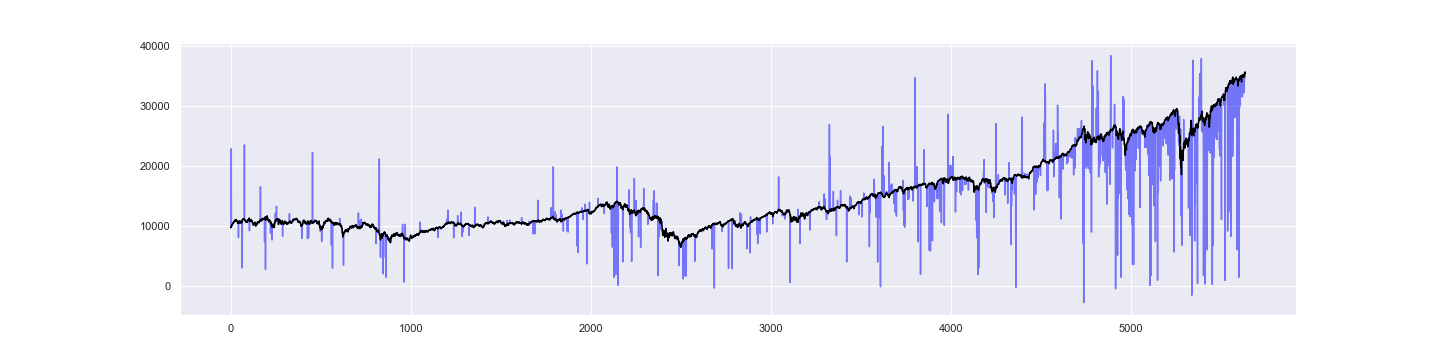}
\caption{Application of the model to Dow Jones Index. Black line indicates Dow Jones index, and blue color line over or under the black line indicates the model's prediction of price change direction.}
\label{fig:mesh2}
\end{figure}

\end{center}

As for the bitcoin market, the model successfully predicted two major crashes, each in 2017 and 2021.

\begin{center}
\begin{figure}[h]
\includegraphics[width=16cm]{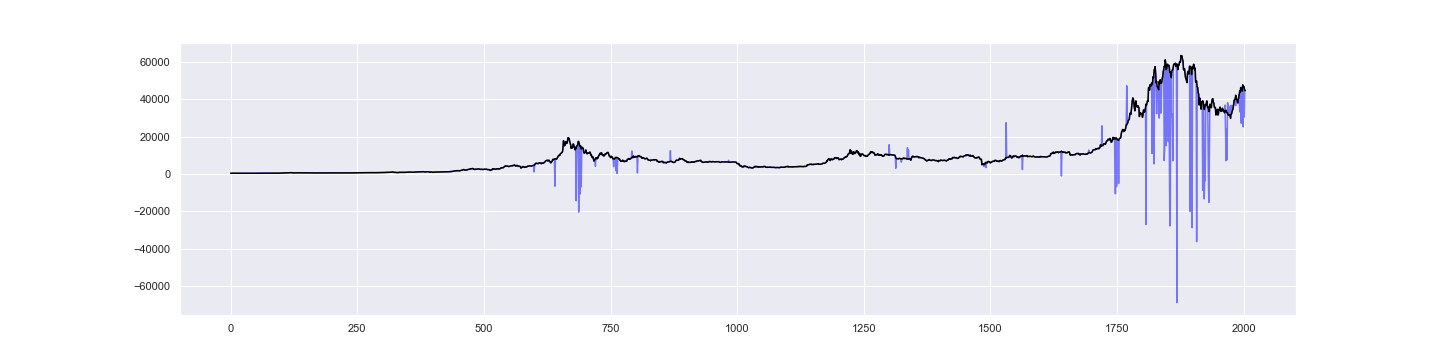}

\caption{Application of the model to Bitcoin makert price. Black line indicates Dow Jones index, and blue color line over or under the black line indicates the model's prediction of price change direction.}
\label{fig:mesh1}
\end{figure}
\end{center}

\section{Conclusion}

While having some implementation issues and prediction errors, the model presented in this paper suggest the logical explanation and prediction for irrational economic behaviors. Although its capability of clear prediction is limited at this point, we expect to improve the model and its implementation methodology and use the model as the potential index for understanding the irrational nature of our economy.

\vskip 0.2in

\bibliography{main}
\bibliographystyle{plain}
\end{document}